\documentclass{jetpl}
\usepackage{bm,amsfonts,amssymb}
\twocolumn

\newcommand{\be}{\begin{equation}}
\newcommand{\ee}{\end{equation}}
\newcommand{\beq}{\begin{eqnarray}}
\newcommand{\eeq}{\end{eqnarray}}
\newcommand{\vex}{{\bm x}}
\newcommand{\ver}{{\bm r}}
\newcommand{\veR}{{\bm R}}
\newcommand{\vep}{{\bm p}}
\newcommand{\veL}{{\bm L}}
\newcommand{\veS}{{\bm S}}
\newcommand{\ven}{{\bm n}}
\newcommand{\vesig}{{\bm \sigma}}

\title{Dynamical suppression of the spin--orbit interaction in hadrons}
\rtitle{Dynamical suppression of the spin--orbit interaction in hadrons}

\author{A.\,M.\,Badalian, A.\,V.\,Nefediev, and Yu.\,A.\,Simonov}
\address{Institute of Theoretical and Experimental Physics,
117218, B.Cheremushkinskaya 25, Moscow, Russia}

\rauthor{A.\,M.\,Badalian, A.\,V.\,Nefediev, and Yu.\,A.\,Simonov}

\abstract{Spin--dependent interaction in hadrons is considered in the approach of the QCD string.
The string moment of inertia, which ensures the correct (inverse) Regge slope $2\pi\sigma$,
is found to suppress the spin--orbit interaction. For light quarks and moderate angular momenta
the suppression constitutes around 25\% whereas for large angular momenta the spin--orbit interaction is
suppressed by the factor $L^{-2/5}$. For heavy quarks the effect manifests itself as a string correction
for the spin-dependent potential.}


\begin{document}
\maketitle

\section{Introduction}

Precise knowledge of the spin--orbital quark-anti\-quark interaction
can be considered as one of the most important goals in the QCD
spectroscopy. It could help one to identify newly discovered orbital
excitations, calculate hadronic shifts for mesons with better
accuracy, to understand effects responsible for a suppression of
the spin--orbit splittings in higher mesons and baryons, and so on. Nevertheless,
some important features of the spin--orbit interaction remain unclear. In  first turn it refers
to the order of levels inside $nP$ multiplets of heavy--light mesons and to small magnitudes of the fine
structure splittings of their higher excitations.
The problem of the $P$-levels ordering in the $D$- and $B$-mesons has been attacked in a number of approaches.
The inversion was predicted, for example, in Refs.~\cite{isg,Faust,KN}.
In the meantime, the absence of the spin--orbit inversion in heavy--light mesons has been
studied in a recent paper \cite{14}, where in a relativistic potential model the
chiral radiative corrections have been taken into account and
shown that such corrections provide the absence of the spin--orbit inversion.
It is obvious that precise knowledge of the nonperturbative spin--dependent
potentials is strongly needed in order to resolve this puzzle.
In particular, it is well known that the spin--orbit interaction in a heavy quarkonium, to the order $O(1/m^2)$,
takes the form \cite{EF,Gr}:
\beq
V_{SO}(r)&=&\left(\frac{{\bm\sigma}_1\veL}{4m_1^2}+\frac{{\bm\sigma}_2\veL}{4m_2^2}\right)
\left(\frac1r\frac{d\varepsilon}{dr}+\frac2r\frac{dV_1}{dr}\right)\nonumber\\[-2mm]
\label{SO0}\\[-2mm]
&+&\frac{({\bm\sigma}_1+{\bm\sigma}_2)\veL}{2m_1m_2}\frac1r\frac{dV_2}{dr},\nonumber
\eeq
with $\varepsilon (r)$ being the static confining interaction and
\be
V_1=-\varepsilon\;\quad V_2=0,
\ee
if confinement is a Lorentz scalar, while, for the time component of the Lorentz vector, the relation reads:
\be
V_1=0\;\quad V_2=\varepsilon.
\ee
Generalisation of Eq.~(\ref{SO0}) to the case of light quarks interacting through the QCD string, rather than by a static potential, is an
important problem. Partially the solution to this problem is known. Indeed, if light quarks interact through a static potential then the
relation (\ref{SO0}) holds, but the current quark masses are to be substituted by the averaged quark energies
\cite{11,Lisbon}, which appear quite naturally in the einbein field formalism (see below) \cite{DKS}.

In this paper we address another problem,
namely, the influence of the dynamics of gluonic degrees of freedom in quarkonia, expressed in terms of the QCD string, on the form of the
spin--orbit interaction.
We use the Field Correlator Method (FCM) \cite{13} to study spin--dependent interactions in light quarkonia. The proper dynamics of
the string between the quark and the antiquark is taken into account, and we show how the string moment of inertia affects the spin--orbit
interaction, namely, how it suppresses the latter.
Generalisation of the results for baryons and for systems with the constituent glue is straightforward.

\section{Spinless heavy--light system and the effect of the QCD string inertia}

In this chapter we give a brief derivation of the spinless Hamiltonian for a meson in the formalism of the QCD string.
Usage of the open bosonic string in hadronic physics has a long history, and a considerable progress was achieved in canonical
quantisation of such a string with a selfconsistent elimination of vibrational modes \cite{Rh}. The straight-line string can be quantised
naturally in terms of the angular momentum \cite{pr}, the einbein field formalism providing a convenient tool for such a 
quantisation \cite{kn1}. Inclusion of spinning quarks at the ends of the open string allowed one to build a manifestly Poincar{\'e} invariant
string quark model and to investigate the spectrum of conventional quark--antiquark mesons \cite{sol}. Since the string degrees of freedom are taken 
into account in this model, the string inertia influences the spectrum, and the Regge trajectories come out with the correct slope. 
In the meantime, this model lacks some spin-dependent potentials which cannot fit into the scheme adopted in Ref.~\cite{sol}. In this paper we stick to
the Hamiltonian description of quark--antiquark mesons and use the Field Correlator Method. In this method the Lagrangian of the string with 
quarks at the end appears naturally due to the properties of the stochastic vacuum gluonic fields, which lead to the area law asymptotic for the 
isolated Wilson loop formed by the quark trajectories \cite{11,DKS}. In addition, in this formalism, the spin-dependent interquark potentials can be 
derived in a selfconsistent way \cite{11,Lisbon}. Thus, following this method, we start from the in-- and out--states of the quark--antiquark meson,
\be
\Psi^{({\rm in, out})}_{q\bar q}(x,y|A)=\bar{\Psi}_{\bar q}(x)\Phi(x,y)\Psi_q(y),
\ee
with gauge invariance guaranteed by the parallel transporter
\be
\Phi(x,y)=P\exp{\left(ig\int_{y}^{x}dz_{\mu}A_{\mu}^at^a\right)}.
\ee
Then we build the Green's function,
\beq
G_{q\bar q}&=&\langle\Psi_{q\bar q}^{({\rm out})}(\bar{x},\bar{y}|A)
\Psi^{({\rm in})\dagger}_{q\bar q}(x,y|A)\rangle_{q\bar{q}A}\nonumber\\[-1mm]
\\[-1mm]
&=&\langle {\rm Tr}S_q(\bar{x},x|A)\Phi(x,y)S_{\bar{q}}(y,\bar{y}|A)\Phi(\bar{y},\bar{x})\rangle_A\nonumber,
\eeq
where $S_{q}$ and $S_{\bar{q}}$ are the propagators of the quark and the antiquark, respectively, in the background gluonic field.
The spin--independent interaction in this quark--antiquark system is described by the Wilson loop,
\be
W(C)=P\exp{\left(ig\oint_Cdz_{\mu}A_{\mu}\right)},
\ee
averaged over the background gluonic field, where the contour $C$ runs over the quark trajectories.
The averaging can be done using the cumulant expansion \cite{13},
\be
\langle Tr W\rangle\propto\exp\left[-\frac12\int_Sds_{\mu\nu}(x)ds_{\lambda\rho}(x')
D_{\mu\nu\lambda\rho}(x-x')\right],
\label{W2}
\ee
where
\be
D_{\mu\nu\lambda\rho}(x-x')=\left\langle\left\langle Tr F_{\mu\nu}(x)\Phi(x,x')F_{\lambda\rho}(x')\Phi(x',x)\right\rangle\right\rangle
\ee
is the bilocal irreducible correlator of gluonic fields, and higher correlators of the form $\langle\langle FF\ldots F\rangle\rangle$ are
neglected (we work in the Gaussian approximation). This correlator reads \cite{13}:
\beq
D_{\mu\nu\lambda\rho}(x-x')&=&(\delta_{\mu\lambda}\delta_{\nu\rho}-\delta_{\mu\rho}\delta_{\nu\lambda})D(x-x'),\label{DDD}\\
+\frac12\left[\frac{\partial}{\partial x_\mu}(x_\lambda\delta_{\nu\rho}\right.&-&\left.x_\rho\delta_{\lambda\nu})+
\genfrac{(}{)}{0pt}{0}{\mu\leftrightarrow\nu}{\lambda\leftrightarrow\rho}\right]D_1(x-x').\nonumber
\eeq
In this paper we concentrate on the first term in Eq.~(\ref{DDD}), which contributes to confinement.
In the meantime, in the final result for the spin--orbit interaction (see Eqs.~(\ref{V0})-(\ref{V2}) below),
we also give the contribution of the second term.
The profile function $D(x-y)$ decreases in all directions of the Euclidean space, and this decrease is governed by the gluonic correlation
length $T_g$. Lattice simulations give rather small values of $T_g$: from $0.15\div 0.3$ fm, in
Ref.~\cite{lattice}, to $0.1$ fm, in Ref.~\cite{km0}.

Then the final result for the spin-independent part reads:
\be
\langle Tr W(C)\rangle\sim \exp{(-\sigma S_{\rm min})},
\ee
with $S_{\rm min}$ being the area of the minimal surface swept by the quark and antiquark trajectories,
\be
S_{\rm min}=\int_0^Tdt\int_0^1d\beta\sqrt{(\dot{w}w')^2-\dot{w}^2w'^2},
\label{Smin}
\ee
where for the profile function of the string $w_\mu(t,\beta)$ we adopt the straight--line ansatz:
\be
w_{\mu}(t,\beta)=\beta x_{1\mu}(t)+(1-\beta)x_{2\mu},
\label{anz}
\ee
$x_{1,2}(t)$ being the four--coordinates of the quarks at the ends of the string.
This approximation is valid at least for not large excitations due to
the fact that hybrid excitations responsible for the string
deformation are decoupled from a meson by the mass gap of order 1 GeV.
The string tension $\sigma$ appears as a double integral from the profile function $D(x-x')=D(x_0-x_0',|\vex-\vex'|)$ \cite{13}:
\be
\sigma=2\int_0^\infty d\nu\int_0^\infty d\lambda D(\nu,\lambda).
\label{sigma}
\ee

We choose to consider the system in the laboratory frame
and also to synchronise the quark times,
\be
x_{10}=x_{20}=t.
\ee
The resulting Lagrangian of the string extracted from the action (\ref{Smin}) reads:
\be
L_{\rm str}=-\sigma r\int_0^1d\beta\sqrt{1-[\ven\times(\beta\dot{\vex}_1+(1-\beta)\dot{\vex}_2)]^2},
\label{L2}
\ee
where $\ver=\vex_1-\vex_2$, $\ven=\ver/r$. This interaction Lagrangian is to be supplied by the quark kinetic terms
$-m_1\sqrt{1-\dot{\vex}_1^2}-m_2\sqrt{1-\dot{\vex}_2^2}$.

A convenient way to treat relativistic kinetic terms is the introduction of the auxiliary (or einbein) fields according to the relation
\cite{ein,ein3}:
\be
m\sqrt{1-\dot{\vex}^2}\to\frac{m^2}{2\mu}+\frac{\mu}{2}-\frac{\mu\dot{\vex}^2}{2},
\ee
where the original form of the Lagrangian is restored as soon as the extremum in the einbein $\mu$ is taken.
Alternatively, the einbein variables $\mu_1$, and $\mu_2$ can appear in the system action via the variable change
when, instead of the proper time $\tau$, the laboratory time $t$ is introduced according to the relation:
$d\tau_i=dt/(2\mu_i)$, $(i=1,2)$ \cite{11,DKS}. The physical meaning of the variables $\mu_i$ as the kinetic energy
of the $i$-th particle is discussed in Refs.~\cite{DKS,12}. Similarly to the quark einbeins $\mu_{1,2}$, a continuous einbein
$\nu(\beta)$ can be introduced in order to get rid of the square root in the string term (\ref{L2}) \cite{DKS}.

Since, in the einbein form of the Lagrangian, kinetic terms have the form of the nonrelativistic energies
(notice that the full set of
relativistic corrections is summed by the procedure of taking extrema in the einbeins), we proceed from the coordinates of the quarks to the
centre-of-mass position and the relative coordinate, as
\beq
\veR=\zeta_1\vex_1+(1-\zeta_1)\vex_2,\quad\ver=\vex_1-\vex_2,\nonumber\\[-1mm]
\label{Rr}\\[-1mm]
\zeta_i=\frac{\mu_i+\int_0^1\beta\nu d\beta}{\mu_1+\mu_2+ \int^1_0 \nu d\beta},\quad i=1,2.\nonumber
\eeq
In terms of these new variables the Lagrangian of the system reads
(we omit the part responsible for the centre-of-mass motion retaining only the relative motion):

$$
L=\frac{m^2_1}{2\mu_1}+\frac{m^2_2}{2\mu_2}+\frac{\mu_1}{2}+\frac{\mu_2}{2}+
\int^1_0\frac{\nu}{2}d\beta+\int^1_0\frac{\sigma^2r^2}{2\nu}d\beta
$$
\be
+\frac12\mu(\dot{\ver}\ven)^2+\frac12\tilde{\mu}[\dot{\ver}\times \ven ]^2,
\label{13}
\ee
where we have defined the reduced masses for the angular and radial motion separately:
\be
\mu=\frac{\mu_1\mu_2}{\mu_1+\mu_2},
\label{murad}
\ee
\be
\tilde{\mu}=\mu_1(1-\zeta_1)^2+\mu_2\zeta_1^2+\int^1_0(\beta-\zeta_1)^2\nu d\beta.
\label{muang}
\ee

In order to proceed from the Lagrangian (\ref{13}) to the Hamiltonian of the system we are to apply the standard procedure
and to define the canonical momentum as
\be
\vep=\frac{\partial L}{\partial \dot{\ver}}=\mu(\ven\dot\ver)\ven+\tilde{\mu}(\dot\ver-\ven(\ven\dot{\ver})),
\label{16}
\ee
with the radial component and transverse components being
\be
(\ven\vep)=\mu(\ven\dot{\ver}),\quad [\ven\times\vep]=\tilde{\mu}[\ven\times\dot{\ver}].
\ee
One can see therefore that the angular momentum is proportional to the
total moment inertia $\tilde{\mu}$ which includes both the effective masses of the quarks and also the proper string moment of inertia
(the last term in $\tilde{\mu}$). Thus we arrive at the spin-independent part of the Hamiltonian \cite{33}:
\be
H=\sum_{i=1}^2\left[\frac{m_i^2}{2\mu_i}+\frac{\mu_i}{2}\right]+\int^1_0d\beta\left[\frac{\sigma^2r^2}{2\nu}+\frac{\nu}{2}\right]
+\frac{p_r^2}{2\mu}+\frac{\veL^2}{2\tilde{\mu}r^2}.
\label{Hm}
\ee
Notice that the difference between $\mu$ and $\tilde{\mu}$ in the last term in
Eq.~(\ref{Hm}) gives rise to the so-called string correction
in the spin-independent Hamiltonian \cite{11,12} and, finally, to the
correct Regge slope $M^2=2\pi\sigma J$ \cite{DKS,MNS}.

An important prediction can be drawn from Eq.~(\ref{Hm}) for radially excited mesons.
Indeed, for $L\neq 0$ and $n_r\gg L$, one has $\tilde{\mu}\gtrsim\mu$ which leads to a suppression of the
angular--momentum--dependent term in the Hamiltonian (\ref{Hm}). As a result, such radially
excited mesons should have masses rather close to those of the $S$-wave levels with the same $n_r$.

The main conclusion of this chapter is that, every time angular momentum appears in the Hamiltonian, it is accompanied by the total inertia
$\tilde{\mu}$ which is larger that the pure quark inertia $\mu$. The physical meaning of this phenomenon is obvious: rotation involves not
only the quarks but also the string, and the inertia of the latter is to be taken into account as well \cite{NST0}.

\section{Nonperturbative spin--orbit interactions}\label{SDint}

The most economical way to include spin-dependent terms into consideration is to extend the
differential in Eq.~(\ref{W2}) and to include the spinor structure ($\tau$ being the proper time)
\cite{Lisbon}:
\be
ds_{\mu\nu}(x)\to d\pi_{\mu\nu}(x)=ds_{\mu\nu}(x)-i\sigma_{\mu\nu}d\tau,
\label{ssig}
\ee
where $\sigma_{\mu\nu}=\frac{1}{4i}(\gamma_\mu\gamma_\nu-\gamma_\nu\gamma_\mu)$. Then the spin--orbit
interaction results from the mixed terms $ds_{\mu\nu}\sigma_{\lambda\sigma}d\tau$:
\be
L_{SO}=\int d s_{\mu\nu}(w)d\tau_1\sigma^{(1)}_{\lambda\rho}D_{\mu\nu\lambda\rho}(w-x_1)+(1\to 2),
\label{26}
\ee
where
\be
ds_{\mu\nu}=\varepsilon^{ab}\partial_a w_\mu(t,\beta)\partial_b w_\nu(t,\beta)dtd\beta,\quad a,b=\{t,\beta\}.
\ee
For the straight-line string ansatz (\ref{anz}) this yields:
\be
ds_{i4}=r_idtd\beta,\quad ds_{ik}=\varepsilon_{ikm}\rho_mdtd\beta,
\label{19a}
\ee
\be
{\bm \rho}=[\ver\times(\beta\dot{\vex}_1+(1-\beta)\dot{\vex}_2)]=(\beta-\zeta_1)[\ver\times\dot{\ver}]=
(\beta-\zeta_1)\frac\veL{\tilde{\mu}}.
\label{a25}
\ee
In addition, as before, we introduce the laboratory time $t$ instead of the proper quark times as
$d\tau_i=\frac{dt}{2\mu_i}$, $i=1,2$.

Finally, by an explicit calculation, one can arrive at the following modification of Eq.~(\ref{SO0})
(the details of the derivation without string moment of inertia can be found in Refs.~\cite{Lisbon,242}; the
full derivation with the string inertia will be given in Ref.~\cite{BNS3}):
\be
\left(\frac{{\bm\sigma}_1\veL}{4m_1^2}+\frac{{\bm\sigma}_2\veL}{4m_2^2}\right)\frac1r\frac{d\varepsilon}{dr}\to
\frac{1}{r}\int_0^\infty d\nu\int_0^r d\lambda \left[\vphantom{\frac12}D+D_1\right.
\label{V0}
\ee
$$
\left.+(\lambda^2+\nu^2)\frac{\partial D_1}{\partial\nu^2}\right]\left[(1-\zeta_1)\frac{{\bm\sigma}_1\veL}{\mu_1\tilde{\mu}}+(1-\zeta_2)\frac{{\bm\sigma}_2\veL}{\mu_2\tilde{\mu}}\right],
$$
$$
\left(\frac{{\bm\sigma}_1\veL}{4m_1^2}+\frac{{\bm\sigma}_2\veL}{4m_2^2}\right)\frac2r\frac{dV_1}{dr}\to-
\frac2r\int^\infty_0d\nu\int^r_0d\lambda D
$$
\be
\times\left(1-\frac{\lambda}{r}\right)\left[(1-\zeta_1)\frac{{\bm\sigma}_1\veL}{\mu_1\tilde{\mu}}+(1-\zeta_2)\frac{{\bm\sigma}_2\veL}{\mu_2\tilde{\mu}}\right],
\label{V1}
\ee
$$
\frac{({\bm\sigma}_1+{\bm\sigma}_2)\veL}{2m_1m_2}\frac1r\frac{dV_2}{dr}\to
\frac{2}{r^2}\int^\infty_0d\nu\int^r_0\lambda d\lambda \left[\vphantom{\frac12}D+D_1\right.
$$
\be
\left.+\lambda^2\frac{\partial D_1}{\partial\lambda^2}\right]\frac{(\vesig_1+\vesig_2)\veL}{\tilde{\mu}}\left(\frac{\zeta_1}{\mu_1}\right),
\label{V2}
\ee
where the contribution of the nonconfining part $D_1$ of the correlator (\ref{DDD}) is quoted for the sake of
completeness.

It is important to stress that this result is not due to the $1/m$ expansion but is obtained with the only
approximation made being the Gaussian approximation for field correlators.
Accuracy of this approximation was checked to be of the order of one percent \cite{30}. For heavy quarks the effective
masses $\mu_{1,2}$
appear to be approximately equal to the current quark masses, and the contribution of the string inertia is negligible. Thus the result
(\ref{SO0}) is recovered. Notice that Gromes relations \cite{Gr} are therefore preserved automatically.
In the meantime, for light quarks, the effective masses $\mu_{1,2}$ are not at all small: they are calculated to be
of order of the interaction scale, that is $\mu_{1,2}\simeq\sqrt{\sigma}\simeq 0.4$ GeV and they grow for excited mesons. This observation
validates the form of the nonperturbative spin--orbit interaction (\ref{V0})-(\ref{V2}) for both heavy and light quarks.

\section{Discussion}

The effect discussed in this paper, namely, the role played by the proper dynamics of the QCD string in the meson, leads to an extra suppression of the spin--orbit
interaction due to the string inertia. Let us estimate such a suppression. To this end, let us check the two limiting cases.
For a heavy--light system with $m_2\to\infty$, one has $\mu_2\to\infty$, $\zeta_1\to 0$, and then the denominator contains the term
\be
\mu_1\left[\mu_1+\int^1_0\beta^2\nu(\beta)d\beta\right],
\label{29}
\ee
which should be confronted with just $\mu_1^2$, if the contribution of the string is neglected.
Another limiting case is given by equal masses. Then $\mu_1=\mu_2=\mu_0$, $\zeta_i=1/2$, and thus,
instead of $\mu_0^2$, we have the denominator:
\be
\mu_0\left[\mu_0+2\int^1_0\left(\beta -\frac12\right)^2\nu(\beta)d\beta\right],
\label{30}
\ee
where the moment of inertia of the string, rotating around its center, is taken into account by the last term
in the square brackets.

This is exactly the effect we are looking for: in both limits considered above the string contributes
to the total moment of inertia of the system and this extra term is always present in the denominator, whenever the angular momentum
$\veL$ appears in the numerator. Due to this modification the spin--orbit interaction becomes weaker than in the ``standard'' case
when the moment of inertia of the rotating string is neglected. Indeed, in the extreme case of highly
orbitally excited states, $L\gg n_r$, the mass $\tilde{\mu}$ is saturated by the string inertia term \cite{33},
so that
\be
V_{SO}\simeq \frac{\mu}{\langle\nu\rangle}V_{SO}^{(0)}\propto\frac{1}{L^{2/5}}V_{SO}^{(0)},
\ee
where $V_{SO}^{(0)}$ is the ``standard" spin--orbit potential with the string inertia neglected,
$\langle\nu\rangle$ stands for the corresponding integral from the extremal value of the einbein
$\nu(\beta)$. In the opposite situation of $L\lesssim n_r$ one has $\langle\nu\rangle\simeq\mu$.
In order to quantify the effect in this case, let us consider an idealised heavy--light
meson ($m_1=0$, $m_2=\infty$) and a simplified version of the
Hamiltonian (\ref{Hm}), with $\tilde{\mu}=\mu$. Then the extremum in the einbein $\nu(\beta)$ can be taken explicitly, with the result
$\nu_{\rm ext}(\beta)=\sigma r$, and we arrive at the well-known Salpeter Hamiltonian with the einbein $\mu_1$ introduced in the kinetic term
of the light quark,
\be
H=\frac{\mu_1}{2}+\frac{\vep^2}{2\mu_1}+\sigma r.
\ee
From the virial theorem it is clear that $\langle\sigma r\rangle=M/2$, with $M$ being the mass of the meson (indeed, if the extremum in $\mu_1$
is taken, the Hamiltonian is symmetric with respect to the kinetic energy and the potential energy interchange, so that they contribute
equally to the total mass of the state). On the other hand, it is easy to
establish that
\be
M(\mu_1)=\frac{\mu_1}{2}+\frac{\sigma^{2/3}}{\mu_1^{1/3}}\varepsilon,
\label{Mmu}
\ee
where $\varepsilon$ is the solution of the nonrelativistic dimensionless Schr{\"o}dinger equation with linear potential. Taking extremum in
$\mu_1$ in Eq.~(\ref{Mmu}) we immediately find that $M=2\mu_1$ and, therefore,
\be
\mu_1=\langle\sigma r\rangle.
\ee
Now, approximating $\nu(\beta)$ in Eq.~(\ref{29}) by $\langle\sigma r\rangle$ we arrive at the denominator change like
\be
\frac{1}{\mu_1^2}\to\frac34\frac{1}{\mu_1^2}.
\label{340}
\ee
It is easy to check that exactly the same result holds for the light--light system. We conclude therefore
that, for light quarks and at low angular momenta, the suppression of the spin--orbit interaction
due to the proper string inertia constitutes around 25\%.
Notice that this suppression has a purely dynamical origin.

For heavy quarks the effect is more moderate. Indeed, for heavy quarks,
$\mu_i\approx m_i$, and the spin--orbit interaction at large interquark separations takes the form
\be
V_{SO}\approx-\frac{\sigma}{mr(m+\frac16\sigma r)}\veS\veL\approx -\frac{\sigma}{m^2r}\veS\veL+\Delta V_{SO},
\ee
where $\veS$ is the total spin of quark--antiquark pair. Here the definition of the string tension (\ref{sigma}) was used, and the einbein $\nu(\beta)$ was substituted by its extremal value
$\sigma r$. The correction to the spin--orbit interaction,
\be
\Delta V_{SO}=\frac{\sigma^2}{6m^3}\veS\veL,
\label{spstr}
\ee
is the string correction for the spin-dependent potential, in analogy to the string correction for the spin-independent potential which
follows from a similar expansion of the last, angular-momentum-dependent term in the Hamiltonian (\ref{Hm}) and which is discussed in
detail in the literature --- see, for example, Refs.~\cite{DKS,33,westr}. For charmonium the string correction (\ref{spstr})
constitutes a few MeV for the lowest states. It grows with the angular momentum and, for excited states,
when the quark mass becomes negligible, it approaches the value given by Eq.~(\ref{340}).

Our results may have some implications for the proton spin physics (for a review see \cite{spn}), and in particular imply a suppression 
of spin-dependent structure functions of baryons for moderate and large $x$ (resonance region).
In the meantime, time spin--spin correlations are not subject to a suppression in our formalism. A vast area of nonperturbative
contributions to spin-dependent effects in parton distributions and Generalised Parton Distributions including small $x$
requires study of spin effects in hybrid baryons. 

Explicit calculations of the mesons spectra with the effect of the string inertia taken
into account are in progress now and will be reported elsewhere. Clearly, generalisation of this effect to
the case of baryons and hadrons with constituent gluons is straightforward, since the einbein field formalism is
applicable for massless particles as well. Finally, the suppression of the spin--orbit interaction predicted
here can be also important for a better understanding of the mixings and mass shifts of the $P$-wave levels
with $J^P=1^+$ in heavy--light mesons, which are very sensitive to the value of the spin--orbit splitting
with respect to the tensor splitting, the latter being not affected by the string rotation \cite{BST}.

This work was supported by the Federal Agen\-cy for Atomic Energy of Russian Fe\-de\-ration, by the Federal Programme of
the Russian Ministry of Industry, Science, and Technology No. 40.052.1.1.1112, and by the grant NSh-4961.2008.2 for the leading scientific
schools. A. M. B. and Yu. A. S. would like to acknowledge the financial support through the grants RFFI-06-02-17012 and RFFI-06-02-17120.
Work of A. N. was supported by RFFI-05-02-04012-NNIOa, DFG-436 RUS 113/820/0-1(R), PTDC/FIS/70843/2006-Fisica,
and by the non-profit ``Dynasty" foundation and ICFPM.

\end{document}